\documentstyle{prhep97}
\input epsf
\def\dofig#1#2{\epsfxsize=#1\centerline{\epsfbox{#2}}}

\def\figfrom#1{\relax}

\def\slashchar#1{\setbox0=\hbox{$#1$}           
   \dimen0=\wd0                                 
   \setbox1=\hbox{/} \dimen1=\wd1               
   \ifdim\dimen0>\dimen1                        
      \rlap{\hbox to \dimen0{\hfil/\hfil}}      
      #1                                        
   \else                                        
      \rlap{\hbox to \dimen1{\hfil$#1$\hfil}}   
      /                                         
   \fi}                                         %

\def\simge{
    \mathrel{\rlap{\raise 0.511ex
        \hbox{$>$}}{\lower 0.511ex \hbox{$\sim$}}}}

\def\simle{
    \mathrel{\rlap{\raise 0.511ex 
        \hbox{$<$}}{\lower 0.511ex \hbox{$\sim$}}}}

\def\etmiss{\slashchar{E}_T}
\def\sgn{\mathop{\rm sgn}}

\def\fbi{{\rm fb}^{-1}}

\def\lsp{\tilde\chi_1^0}
\def\tchi{{\tilde\chi}}

\def\tg{{\tilde g}}
\def\tell{{\tilde\ell}}
\def\ttau{{\tilde\tau}}
\def\tb{{\tilde b}}

\def\Meff{M_{\rm eff}}
\def\mhalf{m_{1/2}}

\def\GeV{{\rm GeV}}
\def\MeV{{\rm MeV}}

\twocolumn 
\newdimen\colwidth \colwidth=\textwidth
\advance\colwidth by -\columnsep
\divide\colwidth by 2

\makeatletter
\let\chapter\hid@chapter
\makeatother

\font\twelvess=cmss10 scaled \magstep1

\begin{document}


\begingroup
\parindent=20pt
\thispagestyle{empty}
\moveleft3.25in\vbox to 8.3in{\hsize=6.4in
\vbox{\vskip-40pt}
{\large
\centerline{\twelvess BROOKHAVEN NATIONAL LABORATORY}
\vskip6pt
\hrule
\vskip1pt
\hrule
\vskip4pt
\hbox to \hsize{December, 1997 \hfil BNL--65029}
\vskip3pt
\hrule
\vskip1pt
\hrule
\vskip3pt

\vskip1.5in
\centerline{\LARGE\bf Search for SUSY at LHC: Precision Measurements}
\vskip.5in
\centerline{\bf Frank E. Paige}
\vskip4pt
\centerline{Physics Department}
\centerline{Brookhaven National Laboratory}
\centerline{Upton, NY 11973 USA}
\vskip1in
\centerline{ABSTRACT}
\vskip8pt\narrower\narrower
	Methods to make precision measurements of SUSY masses and
parameters at the CERN Large Hadron Collider are described.
\vskip1in
	To appear in the {\sl Proceedings of the International
Europhysics Conference on High Energy Physics} (Jerusalem, 1997)
\vskip0pt
}

\vfil

	This manuscript has been authored under contract number
DE-AC02-76CH00016 with the U.S. Department of Energy.  Accordingly,
the U.S.  Government retains a non-exclusive, royalty-free license to
publish or reproduce the published form of this contribution, or allow
others to do so, for U.S. Government purposes.
}

\vfill\eject\endgroup


\def\thepage{\relax}


\authorrunning{Frank E. Paige}
\titlerunning{{\talknumber}: SUSY at LHC}
 

\def\talknumber{1814} 

\title{{\talknumber}: Search for SUSY at LHC: Precision Measurements}
\author{Frank~E.~Paige
(paige@bnl.gov)}
\institute{Brookhaven National Laboratory, Upton, NY 11973 USA}

\maketitle

\begin{abstract}
	Methods to make precision measurements of SUSY masses and
parameters at the CERN Large Hadron Collider are described.
\end{abstract}

\section{Introduction}

	It is quite easy to find signals for SUSY at the
LHC.\cite{SUSY,DPF} But every SUSY event contains two missing
$\lsp$'s, so it is not possible to reconstruct masses directly. A
strategy developed recently\cite{Snow,HPSSY} is to start at the bottom
of the SUSY decay chain and work up it, partially reconstructing
specific final states and using kinematic endpoints to determine
combinations of masses. These are then fit to a model to determine the
SUSY parameters. This paper is limited to discussion of this approach;
search limits and inclusive measurements are discussed by
Abdullin.\cite{Abdullin}

\begin{table}[t]
\caption{Parameters for the five LHC SUGRA points. \label{tbl:points}}
\begin{center}
\begin{tabular}{cccccc}
\hline
& $m_0$ & $\mhalf$ & $A_0$ & $\tan\beta$ & $\sgn{\mu}$\\
& (GeV) & (GeV)   & (GeV)   &             &\\
\hline
1 & 400 & 400 &   0 & \phantom{0}2.0 & $+$\\
2 & 400 & 400 &   0 & 10.0           & $+$\\
3 & 200 & 100 &   0 & \phantom{0}2.0 & $-$\\
4 & 800 & 200 &   0 & 10.0           & $+$\\
5 & 100 & 300 & 300 & \phantom{0}2.1 & $+$\\
\hline
\end{tabular}
\vskip-12pt
\end{center}
\end{table}

\begin{table}[t]
\caption{Representative masses for the five LHC SUGRA points in
Table~\protect\ref{tbl:points}. \label{tbl:masses}}
\begin{center}
\begin{tabular}{cccccc}
\hline
& $M_{\tilde g}$ & $M_{\tilde u_R} $ & $M_{\tilde W_1}$ & $M_{\tilde
e_R}$ & $M_h$\\
 & (GeV) & (GeV) & (GeV) & (GeV) & (GeV)\\
\hline
1 & 1004 & 925 & 325 & 430 & 111\\
2 & 1008 & 933 & 321 & 431 & 125\\
3 & \phantom{0}298 & 313 &  \phantom{0}96 & 207 & \phantom{0}68\\
4 & \phantom{0}582 & 910 & 147 & 805 & 117\\
5 & \phantom{0}767 & 664 & 232 & 157 & 104\\
\hline
\end{tabular}
\vskip-12pt
\end{center}
\end{table}

	The LHCC (LHC Program Committee) selected five points in the
minimal SUGRA model\cite{DPF} for detailed study. The parameters of
this model are $m_0$, the common scalar mass; $\mhalf$, the common
gaugino mass; $A_0$, the common trilinear coupling;
$\tan\beta=v_2/v_1$, the ratio of Higgs vacuum expectation values; and
$\sgn\mu$, the sign of the Higgsino mass. These parameters are listed
in Table~\ref{tbl:points}, and representative masses are listed in
Table~\ref{tbl:masses}.  Point~3 is the ``comparison'' point; LEP
would have already found the light Higgs at this point.  Point~5 is
constructed to give the right cold dark matter. Points~1 and 2 have
heavy masses, while Point~4 has heavy squarks. 

\section{Specific Final States}

	This section describes only a few of the final states that
have been studied. For all of these studies, signal and background
events were generated using ISAJET\cite{ISAJET} or
PYTHIA\cite{PYTHIA}, the response of the detector was simulated, and
an analysis was done to select the signal from the background.

\begin{figure}[t]
\dofig{\colwidth}{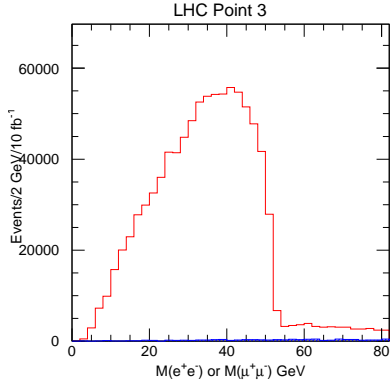}
\caption{Dilepton mass distribution at Point~3 and Standard Model
background (shaded).\protect\cite{HPSSY}\label{c3edge}}
\end{figure}

	$M(\tchi_2^0)-M(\lsp)$: The prototype of precision
measurements\cite{Snow} is based on the decay $\tchi_2^0 \to \lsp
\ell^+ \ell^-$ at Point~3. Point~3 has unusual branching ratios:
\begin{eqnarray*}
B(\tg \to \tb_1 \bar b + {\rm h.c.}) &=& 89\% \\
B(\tb_1 \to \tchi_2^0 b) &=& 86\% \\
B(\tchi_2^0 \to \lsp \ell^+\ell^-) &=& 2 \times 17\%
\end{eqnarray*}
Events were selected with an $\ell^+\ell^-$ pair with
$p_{T,\ell}>10\,\GeV$ and $\eta<2.5$ and at least two jets tagged as
$b$'s with $p_T>15\,\GeV$ and $\eta<2$.  Efficiencies of 60\% for
tagging $b$'s and 90\% for lepton identification were included. No
$\etmiss$ cut was used.  The resulting dilepton mass distribution,
Figure~\ref{c3edge}, has a spectacular edge at the
$M(\tchi_2^0)-M(\lsp)$ endpoint with almost no Standard Model
background. Determining the position of the edge is much easier than
measuring $M_W$ at the Tevatron, and the statistics are huge. The
estimated error for $10\,\fbi$ is $\Delta(M(\tchi_2^0)-M(\lsp)) =
50\,\MeV$. 

\begin{figure}[t]
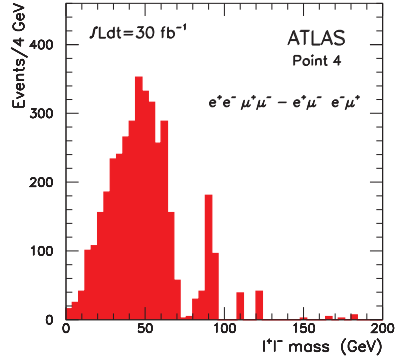

\dofig{\colwidth}{fabiola11.ai}
\caption{Dilepton mass distribution for Point~4.\protect\cite{Fabiola} 
\label{fabiola11}}
\end{figure}

	The low masses and unusual branching ratios make Point~3
particularly easy. But there is a similar edge at Point~4 plus a $Z$
peak coming from decays of the heavier gauginos, as can be seen in
Figure~\ref{fabiola11}.\cite{Fabiola} In this case the estimated error
is $\Delta\left(M(\tchi_2^0)-M(\tchi_1^0)\right) = \pm1\,\GeV$. 
A scan of the SUGRA parameter space\cite{CMS012} finds an observable
signal for $\mhalf \simle 200\,\GeV$ and for a region of small $m_0$
in which the sleptons are light.

\begin{figure}[t]
\dofig{\colwidth}{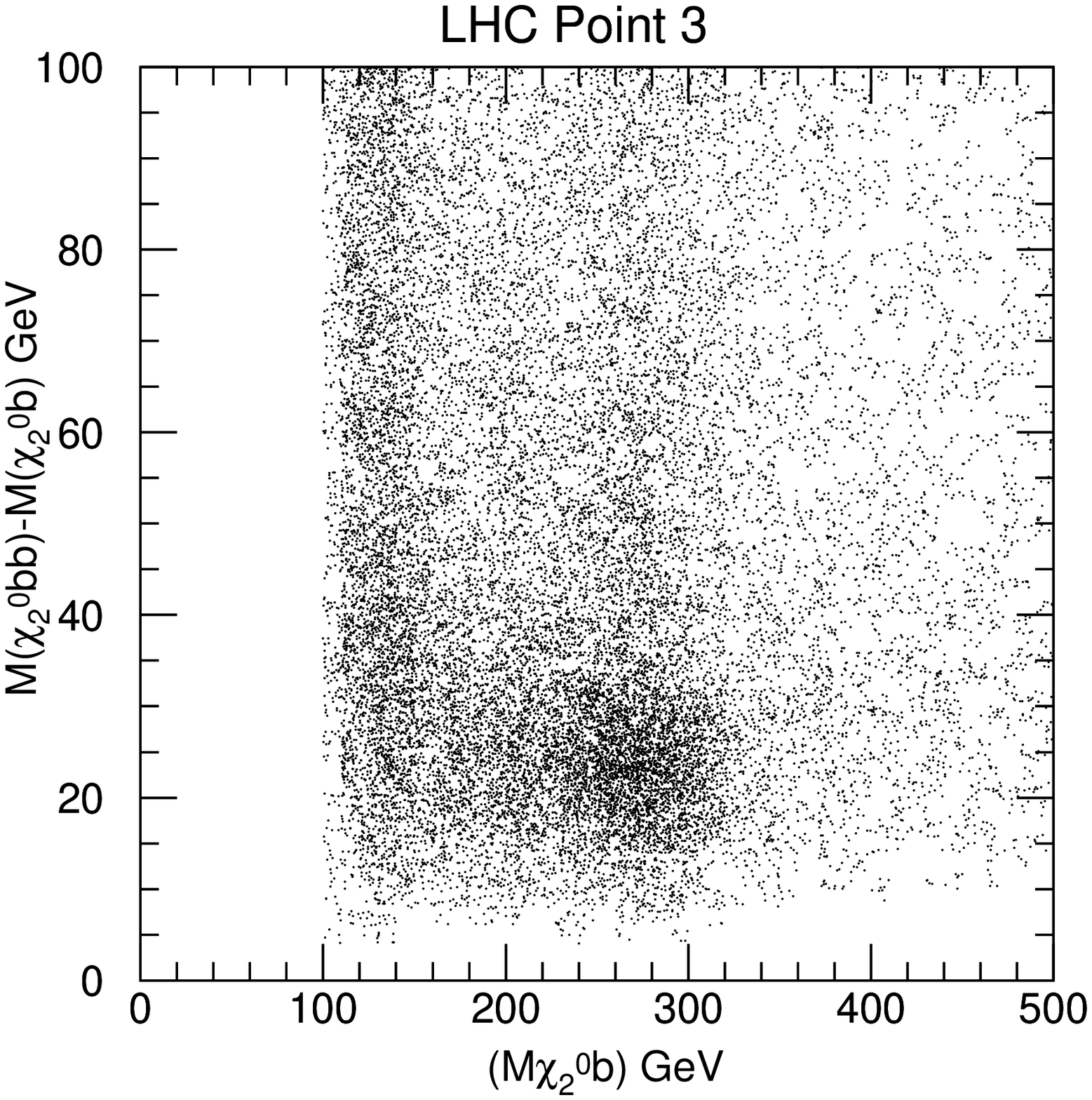}
\caption{Scatter plot of $M(\tg)-M(\tb)$ vs.\ $M(\tb)$.
\protect\cite{HPSSY}\label{c3scatter}}
\end{figure}

	$\tg$ and $\tb_1$: The next step at Point~3 is to combine an
$\ell^+\ell^-$ pair near edge with jets. Events are selected as
before. If the $\ell^+\ell^-$ pair has a mass near the endpoint, then
the $\lsp$ must be soft in $\tchi_2^0$ rest frame, so\hfil
$$
\vec p(\tchi_2^0) \approx \left(1 + {M(\lsp) \over M(\ell\ell)}\right) 
\vec p(\ell\ell) $$ where $M(\lsp)$ must be determined. Lepton pairs
were selected with masses within $10\,\GeV$ of the endpoint and were
combined with one $b$ to make $M(\tb_1)$ and then with a second $b$ to
make $M(\tg)$.  Figure~\ref{c3scatter} shows a scatter plot of all
combinations.  Since the $\bar b$ jet from $\tg \to \tb \bar b$ is
soft, there is good resolution on the $M(\tg)-M(\tb_1)$ mass
difference --- c.f.{} $D^* \to D\pi$. Varying the assumed $\lsp$ mass
gives $\Delta M(\tb_1) = \pm 1.5\Delta M(\lsp) \pm 3\,\GeV$ and
$\Delta(M(\tg) - M(\tb_1)) = \pm 2\,\GeV$.

	$h\to b \bar b$: For Point~5, $\tchi_2^0 \to \lsp h$ is
kinematically allowed. Events are selected with at least four jets
with $p_T>50\,\GeV$, $p_{T,1}>100\,\GeV$, transverse sphericity
$S_T>0.2$, $\Meff = \etmiss + \sum_{i=1}^4\, p_{T,i} > 800\,\GeV$, and
$\etmiss > \max(100\,\GeV, 0.2\Meff)$. Then $M_{bb}$ is plotted for
jets tagged as $b$'s with $p_{T,b}>25\,\GeV$ and $\eta_b<2$.
There is a clear peak with a substantial SUSY background and small
Standard Model background.

\begin{figure}[t]
\dofig{\colwidth}{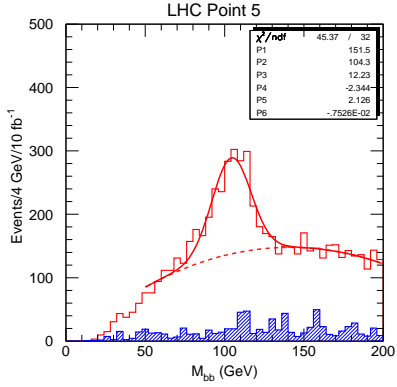}
\caption{$M(bb)$ at Point~5 and Standard Model background
(shaded).\protect\cite{HPSSY} \label{c5mbb}}
\end{figure}

	The two jets from $h \to b \bar b$ can be combined with one of
the two hardest jets in the event to determine the squark mass: the
smaller of the two $b \bar b q$ masses must be less than a function of
the squark mass and the other masses in the decay $\tilde q \to
\tchi_2^0 q \to \lsp h q$.

\begin{figure}[t]
\dofig{\colwidth}{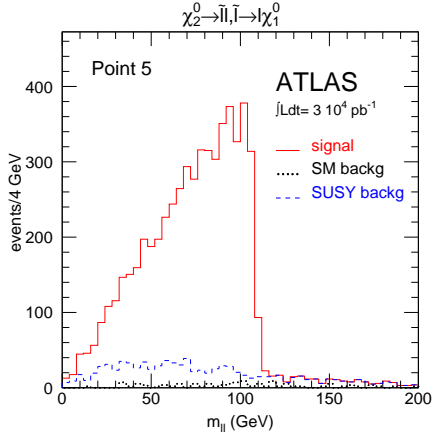}
\caption{$\ell^+\ell^-$ mass distribution for
Point~5.\protect\cite{Polesello} \label{pol16}}
\end{figure}

	$\ell^+\ell^-$ Again: For Point~5 after standard cuts one
finds an edge in Figure~\ref{pol16}\cite{Polesello} for $>M_Z$.  Since
the two-body decay $\tchi_2^0 \to \lsp h$ has been reconstructed at
this point, this edge cannot come from the three-body decay $\tchi_2^0
\to \lsp \ell^+\ell^-$, since the phase space is much smaller. It must
come instead from $\tchi_2^0 \to \tell^\pm \ell^\mp \to \lsp \ell^\pm
\ell^\mp$. Thus the edge determines
$$
M_{\tchi_2^0}\sqrt{1-{M_{\tilde\ell}^2 \over M_{\tilde\chi_2^0}^2}}
\sqrt{1-{M_{\lsp}^2 \over M_{\tilde\ell}^2}}
$$
with an error of $\pm 1\,\GeV$.

	It is possible to have both $\tchi_2^0 \to \tell_R\ell$ and
$\tchi_2^0 \to \lsp \ell\ell$ edges for some choices of the SUGRA
parameters. An example is shown in Figure~\ref{CMSll}.\cite{CMS012}

\begin{figure}[t]
\dofig{\colwidth}{cmsfig7.ai}
\caption{SUGRA $\ell^+\ell^-$ distribution showing edges from both
$\tchi_2^0 \to \tell \ell$ and direct $\tchi_2^0 \to \lsp \ell\ell$
decays.\protect\cite{CMS012} \label{CMSll}}
\end{figure}

	It should in principle be possible to extract the $\tchi_2^0$,
$\tell$, and $\lsp$ masses from a fit to all the dilepton data. This
has not been studied, but as a first step the distribution for the
ratio $p_{T,2}/p_{T,1}$ of lepton $p_T$'s has been examined for $m_0 =
100, 120\,\GeV$. This distribution is clearly exhibits sensitivity to
the slepton mass. The same distribution can also be used to
distinguish two-body and three-body decays.

	$M(\tg)-M(\tchi_2^0), M(\tchi_1^\pm)$: Gluino production
dominates at Point~4.  Previously, an $\ell^+\ell^-$ edge was found at
this point, determining $M(\tchi_2^0)-M(\lsp)$. The strategy for this
analysis is to select
$$
\tg + \tg \to \tchi_2^0 q \bar q +\tchi_1^\pm q \bar q
$$
using leptonic decays to identify $\tchi_2^0$ and $\tchi_1^\pm$ and so
to reduce the combinatorial background.  Then the jet-jet mass should
have a common endpoint since $M(\tchi_2^0) \approx M(\tchi_1^\pm)$.

\begin{figure}[t]
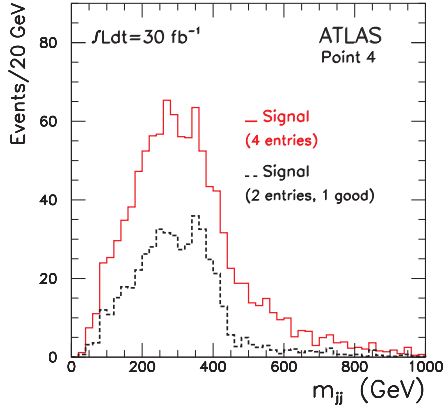

\dofig{\colwidth}{fabiola16.ai}
\caption{Jet-jet mass distribution for Point~4 after cuts described in
the text and corresponding distribution for correct pairing
(dotted).\protect\cite{Fabiola} \label{fabiola16}}
\end{figure}

	The analysis\cite{Fabiola} requires three isolated leptons
with $p_T > 20$, 10, $10\,\GeV$ and$|\eta|<2.5$, one opposite-sign,
same-flavor pair with $M_{\ell\ell}<72\,\GeV$, four jets with $p_T >
150$, 120, 70, $40\,\GeV$, $|\eta|<3.2$, and no additional jets with
$p_T>40\,\GeV$ and $|\eta|<5$ to minimize combinatorics. There are
three pairings per event. The pairing of the two highest and the two
lowest $p_T$ jets is unlikely and is discarded. The distribution for
the remaining pairings, Figure~\ref{fabiola16}, shows an edge at about
the right endpoint.

\section{Fitting SUGRA Parameters}

	Points were generated in SUGRA parameter space, and the masses
were calculated and compared with the combinations of masses
determined by precision measurements.  Fit~I\cite{HPSSY} uses a
smaller set of such measurements, assumes that the Higgs mass can be
related to the SUGRA parameters with an error of $3\,\GeV$, and uses
an integrated luminosity of $10\,\fbi$.  Fit~II\cite{Froid} uses a
larger set of precision measurements plus a few other measurements,
e.g., from changing squark mass and seeing the effect on the highest
$p_T$ jet, assumes a negligible theoretical error on the Higgs mass,
and uses an integrated luminosity of $300\,\fbi$.

\begin{figure}[t]
\dofig{\colwidth}{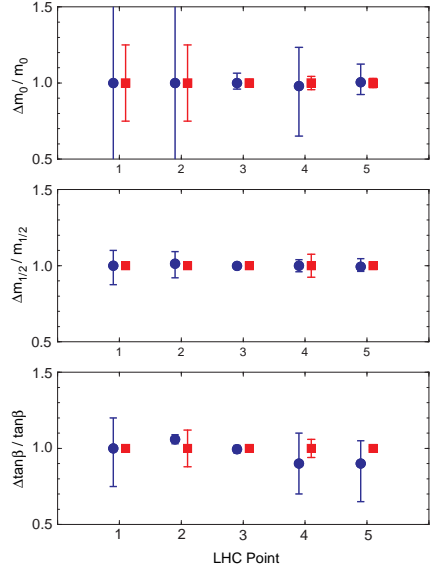}
\caption{Results for Fit~I (circles) and Ultimate Fit~II (squares).
\label{fits}}
\end{figure}

	For both fits the SUGRA parameter space was scanned to
determine the 68\% confidence interval for each parameter. The results
are summarized in Figure~\ref{fits}. Clearly the parameters are quite
well determined. No disconnected regions of parameter space were
found. In particular, $\sgn\mu$ could always be determined.  The gluino
and squark masses are insensitive to $m_0$ at Points~1 and 2, so Fit~I
gives large $m_0$ errors.  Finally, $A_0$ is poorly constrained in all
cases. It is possible to determine the weak scale parameters $A_t$ and
$A_b$, but these are insensitive to $A_0$.

\section{\bf $\tau$ Modes at Large $\tan\beta$}

\begin{figure}[t]
\dofig{\colwidth}{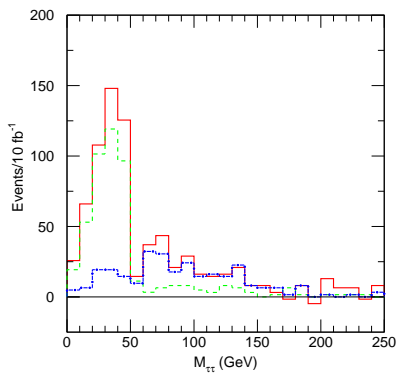}
\caption{Visible $\tau$-$\tau$ mass at a large $\tan\beta$ point and
contributions from $\tchi_2^0$ decays (dashed) had heavy gaugino
decays (dash-dotted).\label{c6mtautau}}
\end{figure}

	For large $\tan\beta$ the $\ttau_1$ can be relatively light.
At the SUGRA point $m_0 = \mhalf = 200\,\GeV$, $A_0 = 0$,
$\tan\beta=45$, $\mu<0$, the decays $\tchi_2^0 \to \ttau_1^\pm
\tau^\mp$ and $\tchi_1^\pm \to \ttau_1^\pm \nu_\tau$ are dominant.
Discovery is still straightforward, but all the analyses discussed in
Section~2 do not apply. One possible approach is to select 3-prong
$\tau$ decays to enhance the visible $\tau$-$\tau$ mass. This is shown
in Figure~\ref{c6mtautau}; it has a clear endpoint at
$M(\tchi_2^0)-M(\lsp)$ plus a continuum from heavier gauginos.  This
example shows that the five LHCC points do not exhaust the
possibilities even of the minimal SUGRA model.

\section{\bf Summary}

	If SUSY exists at electroweak scale, it should be easy to find
signals for it at the LHC. The new result described here is that it is
possible in many cases to make precision measurements of combinations
of SUSY masses, and these measurements can at least in favorable cases
determine the underlying SUSY parameters. While these results are
quite encouraging, it seems likely that some SUSY particles --- including
heavy gauginos, sleptons unless  $\tchi_2^0 \to \tilde \ell \ell$ or
$M(\tilde\ell)\simle200\,\GeV$ to allow substantial Drell-Yan
production, and heavy Higgs bosons --- will be hard to study at the
LHC, so a future lepton-lepton collider could make an important
contribution.

\medskip

	This work was supported in part by the United States
Department of Energy under Contract DE-AC02-76CH00016.

%

\end{document}